\documentstyle{mn} 

\newif\ifAMStwofonts

\ifoldfss
  \ifCUPmtlplainloaded \else
    \NewTextAlphabet{textbfit} {cmbxti10} {}
    \NewTextAlphabet{textbfss} {cmssbx10} {}
    \NewMathAlphabet{mathbfit} {cmbxti10} {} 
    \NewMathAlphabet{mathbfss} {cmssbx10} {} 
  \fi
  \ifAMStwofonts
    \ifCUPmtlplainloaded \else
      \NewSymbolFont{upmath} {eurm10}
      \NewSymbolFont{AMSa} {msam10}
      \NewMathSymbol{\upi}     {0}{upmath}{19}
      \NewMathSymbol{\umu}     {0}{upmath}{16}
      \NewMathSymbol{\upartial}{0}{upmath}{40}
      \NewMathSymbol{\leqslant}{3}{AMSa}{36}
      \NewMathSymbol{\geqslant}{3}{AMSa}{3E}

    \fi
  \fi
\fi 

\ifnfssone
  \newmathalphabet{\mathit}
  \addtoversion{normal}{\mathit}{cmr}{m}{it}
  \addtoversion{bold}{\mathit}{cmr}{bx}{it}
  \newmathalphabet{\mathbfit} 
  \addtoversion{normal}{\mathbfit}{cmr}{bx}{it}
  \addtoversion{bold}{\mathbfit}{cmr}{bx}{it}
  \newmathalphabet{\mathbfss} 
  \addtoversion{normal}{\mathbfss}{cmss}{bx}{n}
  \addtoversion{bold}{\mathbfss}{cmss}{bx}{n}
  \ifAMStwofonts
    \ifCUPmtlplainloaded \else
      \UseAMStwoboldmath
      \makeatletter
      \new@mathgroup\upmath@group
      \define@mathgroup\mv@normal\upmath@group{eur}{m}{n}
      \define@mathgroup\mv@bold\upmath@group{eur}{b}{n}
      \edef\UPM{\hexnumber\upmath@group}
      \new@mathgroup\amsa@group
      \define@mathgroup\mv@normal\amsa@group{msa}{m}{n}
      \define@mathgroup\mv@bold\amsa@group{msa}{m}{n}
      \edef\AMSa{\hexnumber\amsa@group}
      \makeatother
      \mathchardef\upi="0\UPM19
      \mathchardef\umu="0\UPM16
      \mathchardef\upartial="0\UPM40
      \mathchardef\leqslant="3\AMSa36
      \mathchardef\geqslant="3\AMSa3E
    \fi
  \fi
\fi 

\ifnfsstwo
  \DeclareMathAlphabet{\mathbfit}{OT1}{cmr}{bx}{it}
  \SetMathAlphabet\mathbfit{bold}{OT1}{cmr}{bx}{it}
  \DeclareMathAlphabet{\mathbfss}{OT1}{cmss}{bx}{n}
  \SetMathAlphabet\mathbfss{bold}{OT1}{cmss}{bx}{n}
  \ifAMStwofonts
    \ifCUPmtlplainloaded \else
      \DeclareSymbolFont{UPM}{U}{eur}{m}{n}
      \SetSymbolFont{UPM}{bold}{U}{eur}{b}{n}
      \DeclareSymbolFont{AMSa}{U}{msa}{m}{n}
      \DeclareMathSymbol{\upi}{0}{UPM}{"19}
      \DeclareMathSymbol{\umu}{0}{UPM}{"16}
      \DeclareMathSymbol{\upartial}{0}{UPM}{"40}
      \DeclareMathSymbol{\leqslant}{3}{AMSa}{"36}
      \DeclareMathSymbol{\geqslant}{3}{AMSa}{"3E}
    \fi
  \fi
\fi 

\ifCUPmtlplainloaded \else
  \ifAMStwofonts \else 
    \def\upi{\pi}
    \def\umu{\mu}
    \def\upartial{\partial}
  \fi
\fi

\title{$UBVI$ CCD photometry of two old open clusters NGC 1798 and NGC 2192}
\author[Hong Soo Park and Myung Gyoon Lee]
{Hong Soo Park and Myung Gyoon Lee\thanks{corresponding author, E-mail: mglee@astrog.snu.ac.kr } \\
Department of Astronomy, Seoul National University, Seoul 151-742, Korea \\
}
\date{Accepted 1998 ??.
      Received 1998 ??;
      in original form 1998 July ??}

\pagerange{\pageref{firstpage}--\pageref{lastpage}}
\pubyear{1998}

\begin{document}

\maketitle

\label{firstpage}

\begin{abstract}
We present $UBVI$ CCD photometry of two open clusters NGC 1798 and NGC 2192
which were little studied before.
Color-magnitude diagrams of these clusters show several features 
typical for old open clusters: a well-defined main-sequence,
a red giant clump, and a small number of red giants. 
The main sequence of NGC 1798 shows a distinct gap at $V \approx 16.2$ mag.
From the surface number density distribution we have measured 
the size of the clusters, obtaining  $8'.3$ (= 10.2 pc) for NGC 1798 
and $7'.3$ (= 7.5 pc) for NGC 2192. 
Then we have determined the reddening, metallicity, and
distance of these clusters using the color-color diagrams and color-magnitude diagrams: $E(B-V)=0.51\pm0.04$, [Fe/H] = $-0.47\pm0.15$ dex and    
 $(m-M)_0=13.1\pm0.2$ ($d=4.2\pm 0.3$ kpc) for NGC 1798, and
$E(B-V)=0.20\pm0.03$, [Fe/H] = $-0.31\pm0.15$ dex and $(m-M)_0=12.7\pm0.2$
 ($d=3.5\pm 0.3$ kpc) for NGC 2192.
The ages of these clusters have been estimated 
using the morphological age indicators and the isochrone fitting
with the Padova isochrones: $1.4 \pm 0.2$ Gyrs for NGC 1798 and
$1.1 \pm 0.1$ Gyrs for NGC 2192.
The luminosity functions of the main sequence stars in these clusters are
found to be similar to other old open clusters.
The metallicity and distance of these clusters are consistent
 with the relation between the metallicity and galactocentric 
distance of other old open clusters. 
\end{abstract}


\section{Introduction}
Old open clusters provide us with an important information for understanding
the early evolution of the Galactic disk. There are about 70 known old open
clusters with age $> 1$ Gyrs \cite{fri95}. 
These clusters are faint in general so that there were few studies about
these clusters until recently. 
With the advent of CCD camera in astronomy, the number of studies
on these clusters has been increasing. However, there are still a significant
number of old open clusters for which basic parameters are not well known.
For example, metallicity is not yet known for about 30 clusters among them.

Recently Phelps et al.\ (1994) and Janes \& Phelps (1994) 
presented an extensive CCD photometric survey of potential old open clusters, 
the results of which were used in the study on the development of the Galactic
disk by Janes \& Phelps (1994). 
In the sample of the clusters studied by Phelps et al.\ there are several 
clusters for which only the non-calibrated photometry is available.

We have chosen two clusters among them, NGC 1798 and NGC 2192, to study the
characteristics of these clusters using $UBVI$ CCD photometry. These clusters
are located in the direction of anti-galactic centre.
To date there is published only one photometric study of these clusters, 
which was given by Phelps et al.\ (1994) who presented non-calibrated $BV$ CCD 
photometry of these clusters.  
From the instrumental color-magnitude diagrams of these clusters 
Phelps et al.\ estimated the ages of these clusters 
using the morphological age indicators, 
obtaining the values of 1.5 Gyrs for NGC 1798 and 1.1 Gyrs for NGC 2192.
However, no other properties of these clusters are yet known.

In this paper we present a study of NGC 1798 and NGC 2192
based on $UBVI$ CCD photometry. We have estimated the basic parameters of
these clusters: size, reddening, metallicity, distance, and age. 
Also we have derived the luminosity function of the main sequence 
stars in these clusters.
Section 2 describes the observations and data reduction.
Sections 3 and 4 present the analysis for NGC 1798 and NGC 2192, respectively.
Section 5 discusses the results. Finally Section 6 summarizes the primary
results.

\section[]{OBSERVATIONS AND DATA REDUCTION}

\subsection{Observations}
$UBVI$ CCD images of NGC 1798 and NGC 2192 were obtained using the 
Photometrics 512 CCD camera
at the Sobaeksan Observatory 61cm telescope in Korea for several observing
runs between 1996 November and 1997 October. We have used also $BV$ CCD images
of the central region of NGC 1798 obtained by Chul Hee Kim using the Tek 1024
CCD camera at the Vainu Bappu Observatory 2.3m telescope in India on March 4, 1998.
The observing log is given in Table 1.

The original CCD images were flattened after bias subtraction and several 
exposures for each filter were combined into a single image for further
reduction. The sizes of the field in a CCD image are
$4'.3 \times 4'.3$ for the PM 512 CCD image, and
$10'.6 \times 10'.6$ for the Tek 1024 CCD image.
The gain and readout noise are, respectively, 
9 electrons/ADU and 10.4 electrons for the PM 512 CCD, and
9 electrons/ADU and  10.4 electrons for the Tek 1024 CCD.

Figs. 1 and 2 illustrate grey scale maps of the $V$ CCD images of NGC 1798 
and NGC 2192 made by mosaicing the images of the observed regions.
It is seen from these figures that NGC 1798 is a relatively rich  open cluster,
while NGC 2192 is a relatively poor open cluster.

\subsection{Data Reduction}

Instrumental magnitudes of the stars in the CCD images were obtained
using the digital stellar photometry reduction program IRAF\footnote[1]{IRAF is distributed by the National Optical Astronomy Observatories, which are 
operated by the Association of  Universities for Research in Astronomy,
Inc. under contract with the National Science Foundation.}/DAOPHOT
(Stetson 1987, Davis 1994). The resulting instrumental magnitudes were
transformed onto the standard system using the standard stars from Landolt
(1992) and the M67 stars in Montgomery et al.\ (1993) observed on 
the same photometric nights.  The transformation equations are 

\[ V= v + a_V (b-v) + k_V X + Z_V, \]
\[ (B-V)= a_{BV} (b-v) + k_{BV} X + Z_{BV}, \]
\[(U-B)= a_{UB} (u-b) + k_{UB} X + Z_{UB}, {\rm and} \]
\[(V-I)= a_{VI} (v-i) + k_{VI} X + Z_{VI},\]
where the lower case symbols represent instrumental magnitudes
derived from the CCD images and the upper case symbols represent the
standard system values. X is the airmass at the midpoint of the observations.
The results of the transformation are summarized in Table 2.
The data obtained on non-photometric nights
were calibrated using the photometric data for the overlapped region.

The total number of the measured stars is 1,416 for NGC 1798 and
409 for NGC 2192. Tables 3 and 4 list the photometry of the bright stars
in the C-regions of NGC 1798 and NGC 2192, respectively.
The $X$ and $Y$ coordinates listed in Table 3 and 4 are given in units of
CCD pixel ($= 0''.50$). The $X$ and $Y$ values are increasing toward north
and west, respectively.

We have divided the entire region of the fields into several regions, as shown
in Figs. 1 and 2, for the analysis of the data.
The C-region represents the central region of the cluster, and
the F-regions (F, Fb, Fir, and Fi regions) represent the control field regions,
and the N-region represents the intermediate region between the central
region and the field region.
The radius of the C-region is 300 pixel for NGC 1798 and NGC 2192.
The ratio of the areas of the C-region, N-region, Fb-region, and
(Fi + Fir)-regions for NGC 1798 is 1:1.50:1.00:1.07, and
the ratio of the areas of the C-region, N-region, and F-region for NGC 2192
is 1:1.26:0.98.

\section{ANALYSIS FOR NGC 1798}

\subsection{The Size of the Cluster}

We have investigated the structure of NGC 1798 using starcounts.
The centre of the cluster is estimated to be at the position of
($X = 710$ pixel, $Y=1110$ pixel), using the centroid method.
Fig. 3 illustrates the projected surface number density profile derived
from counting stars with $V<19.5$ mag in the entire CCD field. 
The magnitude cutoff for starcounts was set so that the counts should be 
free of any photometric incompleteness problem.
Fig. 3 shows that most of the stars in NGC 1798 are concentrated 
within the radius of 250 pixel ($= 125''$), 
and that the outskirts of the cluster extend out to about 500 pixel 
($=250''$) from the center. 
The number density changes little with radius beyond 500 pixel, showing that
the outer region of the observed field can be used as a control field.
Therefore we have estimated the approximate size of NGC 1798 for which the cluster 
blends in with the field to be about $500''$
in diameter, which corresponds to a linear size of 10.2 pc 
for the distance of NGC 1798 as determined below.

\subsection{Color-Magnitude Diagrams}

Figs. 4 and 5 show the $V-(B-V)$ and $V-(V-I)$ color-magnitude diagrams (CMDs)
of the measured stars in the observed regions in NGC 1798.
These figures show that the C-region consists mostly of the members of NGC 1798 with some contamination of the field stars, while the F-regions consist
mostly of the field stars. The N-region is intermediate between the
C-region and the F-region.

The distinguishable features seen in the color-magnitude diagrams of 
the C-region are:
(a) There is a well-defined main sequence the top of which is located 
at $V \approx 16$ mag;
(b) There is seen a distinct gap at $V \approx 16.2$ mag in the main sequence,
which is often seen in other old open clusters (e.g. M67);
(c) There is a poorly defined red giant branch and these is seen some excess of
stars around $(B-V)=1.3$ and $V=15.6$ mag on this giant branch,  
which is remarked by the small box in the figures. This may be a random
excess of stars. However, the positions of the stars in the CMDs are consistent with the
positions of known red giant clump in other old open clusters. 
Therefore most of these stars are probably red giant clump stars; and
(d) There are a small number of stars along the locus of the red giant
branch.

\subsection{Reddening and Metallicity}

NGC 1798 is located close to the galactic plane in the anti-galactic
centre direction ($b=4^\circ.85$ and $l=160^\circ.76$) so that 
it is expected that the reddening toward this cluster is significant.
We have estimated the reddening for NGC 1798 using two methods as follows.

First we have used the mean color of the red giant clump. 
Janes \& Phelps (1994) estimated the mean color and magnitude 
of the red giant clump in old open clusters 
to be $(B-V)_{RGC}=0.87 \pm 0.02$ and $M_{V, RGC} = 0.59 \pm 0.09$,
when the difference between the red giant clump and the main sequence
turn-off of the clusters, $\delta V$, is smaller than one.
The mean  color of the red giant clump in the C-region is estimated 
to be $(B-V)_{RGC}=1.34 \pm 0.01$ ($(V-I)_{RGC}=1.47 \pm 0.01$, and
$(U-B)_{RGC}=1.62 \pm 0.04$), and the corresponding mean magnitude is
$V_{RGC} = 15.57 \pm 0.05$. $\delta V$ is estimated to be $0.8\pm 0.2$,
which is the same value derived by Phelps et al.\ (1994).
From these data we have derived a value of the reddening, 
$E(B-V) = 0.47 \pm 0.02$.

Secondly we have used the color-color diagram to estimate the reddening
and the metallicity simultaneously.
We have fitted the mean colors of the stars in the C-region with the
color-color relation used in the Padova isochrones \cite{ber94}.
This process requires iteration, because we need to know the age of the
cluster as well as the reddening and metallicity.
We have iterated this process until all three parameters are stabilized.

Fig. 6 illustrates the results of fitting in the $(U-B)-(B-V)$ 
color-color diagram.
It is shown in this figure that the stars in NGC 1798 are reasonably fitted
by the color-color relation of the isochrones for [Fe/H] $= -0.47 \pm 0.15$ 

with a reddening value of $E(B-V)=0.55 \pm 0.05$. 
The error for the metallicity, 0.15, was estimated by comparing isochrones with 
different metallicities. As a reference
the mean locus of the giants for solar abundance given by Schmidt-Kaler (1982)
is also plotted in Fig. 6.
Finally we derive a mean value of the two estimates for the reddening, 
$E(B-V) = 0.51 \pm 0.04$.

\subsection{Distance}

We have estimated the distance to NGC 1798 using two methods as follows.
First we have used the mean magnitude of the red giant clump. We have derived
a value of the apparent distance modulus $(m-M)_V = 14.98 \pm 0.10$ 
from the values
for the mean magnitudes of the red giant clump stars described above.
 
Secondly we have used the the zero-age main sequence
(ZAMS) fitting, following the method described in VandenBerg \& Poll (1989).
VandenBerg \& Poll (1989) presented the semi-empirical ZAMS as a function
of the metallicity [Fe/H] and the helium abundance Y:

\[ V= M_V (B-V)+\delta M_V (Y)+\delta M_V ({\rm [Fe/H]}) \]

\noindent where $\delta M_V (Y) = 2.6 (Y -0.27)$ and
$\delta M_V({\rm [Fe/H]}) = - {\rm [Fe/H]} (1.444 + 0.362 {\rm [Fe/H]})$.

Before the ZAMS fitting, we subtracted statistically the contribution due to
the field stars in the CMDs of the C-region using the CMDs of the Fb-region
for $BV$ photometry and the CMDs of the Fi+Fir region for $VI$ photometry.
The size of the bin used for the subtraction is $\Delta V = 0.25$ and
$\Delta (B-V) = 0.1$. The resulting CMDs are displayed in Fig. 7. 
We used the metallicity of [Fe/H] = --0.47 as derived above and adopted
$Y=0.28$ which is the mean value for old open clusters \cite{gra82}. 
Using this method we have obtained a value of the apparent distance modulus
$(m-M)_V = 14.5 \pm 0.2$. 
Finally we calculate a mean value of the two estimates,
$(m-M)_V = 14.7 \pm 0.2$.
Adopting the extinction law of $A_V = 3.2 E(B-V)$, we derive a value of
the intrinsic distance modulus $(m-M)_0 = 13.1 \pm 0.2$.
This corresponds to a distance of $d=4.2\pm 0.3$ kpc. 

\subsection{Age}

We have estimated the age of NGC 1798 using two methods as follows.
First we have used the morphological age index (MAI) as described in 
Phelps et al.\ (1994). 
Phelps et al.\ (1994) and Janes \& Phelps (1994) presented the MAI--$\delta V$ relation,
\[ MAI[{\rm Gyrs}] = 0.73 \times 10^{(0.256 \delta V + 0.0662 \delta V^2)}.\]
From the value of $\delta V$ derived above, $0.8\pm0.2$ mag, we obtain a value
for the age, MAI $= 1.3\pm0.2$ Gyrs. 

Secondly we have estimated the age of the cluster using  the theoretical
isochrones given by the Padova group \cite{ber94}.
Fitting the isochrones for [Fe/H] = --0.47 to the CMDs of NGC 1798, as shown
in Fig. 8, we estimate the age to be $1.4\pm0.2$ Gyrs.
Both results agree very well.

\subsection{Luminosity Function}

We have derived the $V$ luminosity functions of the main sequence stars
in NGC 1798, which are displayed in Fig. 9.
The Fb-region was used for subtraction of the field star contribution
from the C-region and the magnitude bin size used is 0.5 mag. This control field
may not be far enough from the cluster to derive the field star contribution.
If so, we might have oversubtracted the field contribution, obtaining flatter
luminosity functions than true luminosity functions.
However, the fraction of the cluster members in this field must be, if any, very low,
because the surface number density of this region is almost constant with the radius
as shown in Fig. 3.
The luminosity function of the C-region in Fig. 9(a) increases rapidly up 
to $V\approx 16.5$ mag, and stays almost flat for $V>16.5$ mag.
The luminosity functions of the N-region and the (R+Fir)-region are
steeper than that of the C-region.
A remarkable drop is seen at $V=16.2$ mag ($M_V = 1.5$ mag) 
in the luminosity function of the C-region 
based on smaller bin size of 0.2 mag in Fig.9(b).
This corresponds to the main sequence gap described above.


\section{ANALYSIS FOR NGC 2192}

\subsection{The Size of the Cluster}

We have investigated the structure of NGC 2192 using starcounts.
We could not use the centroid method to estimate the centre of this cluster,
because this cluster is too sparse. So we have used eye-estimate to determine 
the centre of the cluster to be at the position of ($X = 465$ pixel, $Y=930$ pixel).
Fig. 10 illustrates the projected surface number density profile derived
from counting stars with $V<18$ mag in the entire CCD field. 
The magnitude cutoff for starcounts was set so that the counts should be 
free of any photometric incompleteness problem.
Fig. 10 shows that most of the stars in NGC 2192 are concentrated 
within the radius of 200 pixel ($= 100''$), 
and that the outskirts of the cluster extend out to about 440 pixel 
($=220''$) from the centre. 
Therefore the approximate size of NGC 2192 is estimated to be about $440''$
in diameter,which corresponds to a linear size of 7.5 pc 
for the distance of NGC 2192 as determined below.

\subsection{Color-Magnitude Diagrams}

Figs. 11 and 12 show the $V-(B-V)$ and $V-(V-I)$ color-magnitude diagrams
of the measured stars in the observed regions in NGC 2192.
The distinguishable features seen in the color-magnitude diagrams of 
the C-region are:
(a) There is a well-defined main sequence the top of which is located 
at $V \approx 14$ mag;
(b) There are a group of red giant clump stars at $(B-V)=1.1$ and $V=14.2$ mag,
 which are remarked by the small box in the figures; and
(c) There are a small number of stars along the locus of the red giant
branch.

\subsection{Reddening and Metallicity}

NGC 2192 is located 11 degrees above the galactic plane
in the anti-galactic
centre direction ($b=10^\circ.64$ and $l=173^\circ.41$) but higher than
NGC 1798 so that 
it is expected that the reddening toward this cluster is significant but 
smaller than that of NGC 1798.
We have estimated the reddening for NGC 2192 using two methods as applied
for NGC 1798.

First we have used the mean color of the red giant clump. 
The mean  color of the red giant clump in the C-region is estimated 
to be $(B-V)_{RGC}=1.08 \pm 0.01$ ($(V-I)_{RGC}=1.07 \pm 0.01$, and
$(U-B)_{RGC}=0.61\pm0.02$), and the corresponding mean magnitude is
$V_{RGC} = 14.20 \pm 0.05$. $\delta V$ is estimated to be $0.6\pm 0.2$,
which is similar to the value derived by Phelps et al.\ (1994).
From these data we have derived a value of the reddening, 
$E(B-V) = 0.19 \pm 0.03$.

Secondly we have used the color-color diagram to estimate the reddening
and the metallicity simultaneously.
We have fitted the mean colors of the stars in the C-region with the
color-color relation used in the Padova isochrones \cite{ber94}.
Fig. 13 illustrates the results of fitting in the $(U-B)-(B-V)$ 
color-color diagram.
It is shown in this figure that the stars in NGC 2192 are reasonably fitted
by the color-color relation of the isochrones for [Fe/H] $= -0.31 \pm 0.15$ dex 
with a reddening value of $E(B-V)=0.21 \pm 0.01$. 
The error for the metallicity, 0.15, was estimated by comparing isochrones with different
metallicities. As a reference
the mean locus of the giant for solar abundance given by Schmidt-Kaler
is also plotted in Fig. 13.
Finally we derive a mean value of the two estimates for the reddening, 
$E(B-V) = 0.20 \pm 0.03$.

\subsection{Distance}

We have estimated the distance to NGC 2192 using two methods as for NGC 1798.
First we have used the mean magnitude of the red giant clump. We have derived
a value of the apparent distance modulus $(m-M)_V = 13.61 \pm 0.10$ 
from the values
for the mean magnitudes of the red giant clump stars described previously.
 
Secondly we have used the ZAMS fitting. 
Before the ZAMS fitting, we subtracted statistically the contribution due to
the field stars in the CMDs of the C-region using the CMDs of the F-region.
The size of the bin used for the subtraction is $\Delta V = 0.25$ and
$\Delta (B-V) = 0.1$. 

The resulting CMDs are displayed in Fig. 14. 
We used the metallicity of [Fe/H] = --0.31 as derived before and adopted
$Y=0.28$. 
Using this method we have obtained a value of the apparent distance modulus
$(m-M)_V = 13.1 \pm 0.2$. 
Finally we calculate a mean value of the two estimates,
$(m-M)_V = 13.3 \pm 0.2$.
Adopting the extinction law of $A_V = 3.2 E(B-V)$, we derive a value of
the intrinsic distance modulus $(m-M)_0 = 12.7 \pm 0.2$.
This corresponds to a distance of $d=3.5\pm 0.3$ kpc. 

\subsection{Age}

We have estimated the age of NGC 2192 using two methods as follows.
First we have used the morphological age index. 
From the value of $\delta V$ derived above, $0.6\pm0.2$ mag, we obtain a value
for the age, MAI $= 1.1\pm0.2$ Gyrs. 
Secondly we have estimated the age of the cluster using  the theoretical
isochrones given by the Padova group \cite{ber94}.
Fitting the isochrones for [Fe/H] = --0.31 to the CMDs of NGC 2192, as shown
in Fig. 15, we estimate the age to be $1.1\pm0.1$ Gyrs.
Both results agree very well.

\subsection{Luminosity Function}

We have derived the $V$ luminosity functions of the main sequence stars
in NGC 2192, which are displayed in Fig. 16.
The F-region was used for subtraction of the field star contribution
from the C-region.
The luminosity function of the C-region in Fig. 16(a) increases rapidly up 
to $V\approx 14$ mag, and stays almost flat for $V>15$ mag.
The luminosity function of the N-region is
steeper than that of the C-region.
Fig. 16(b) displays a comparison of the luminosity functions of
NGC 1798, NGC 2192, and NGC 7789 which is another old open cluster 
of similar age \cite{rog94}. 
Fig. 16(b) shows that the luminosity functions of these clusters are similar 
in that they are almost flat in the faint part. 
The flattening of the faint part of the luminosity functions of old open clusters
has been known since long, and is believed to be due to evaporation of low mass
stars \cite{fri95}. 

\section{DISCUSSION}

We have determined the metallicity and distance of NGC 1798 and NGC 2192
in this study. We compare them with those of other old open clusters here.
Fig. 17 illustrates the radial metallicity gradient of the old open clusters
compiled by Friel (1995) and supplemented by the data in Wee \& Lee (1996)
and Lee (1997). 
Fig. 17 shows that the mean metallicity decreases as the galactocentric distance
increases. 
The positions of NGC 1798 and NGC 2192 we have obtained in this study
are consistent with the mean trend of the other old open clusters.
The slope we have determined for the entire sample 
including these two clusters is
$\Delta {\rm [Fe/H]} / R_{GC} = -0.086\pm0.011$ dex/kpc, very similar
to that given in Friel (1995), 
$\Delta {\rm [Fe/H]} / R_{GC} = -0.091\pm0.014$ dex/kpc. 

There are only four old open clusters located beyond $R_{GC} = 13$ kpc
in Fig. 17. These four clusters follow the mean trend of decreasing outward. 
However, the number of the clusters is not large enough to decide whether
the metallicty keeps decreasing outward or it stops decreasing somewhere 
beyond $R_{GC} = 13$ kpc and stays constant.
Further studies of more old open clusters 
beyond $R_{GC} = 13$ kpc are needed to investigate this point.

\section{SUMMARY AND CONCLUSIONS}

We have presented $UBVI$ photometry of old open clusters NGC 1798
and NGC 2192. 
From the photometric data we have determined the size, reddening, 
metallicity, distance, and age of these clusters.
The luminosity functions of the main sequence stars in these clusters
are similar to those of the other old open clusters.
The basic properties of these clusters we have determined in this study 
are summarized in Table 5.

\section*{Acknowledgments}
Prof.Chul Hee Kim is thanked for providing the $BV$ CCD images of NGC 1798.
This research is supported in part by
the Korea Science and Engineering Foundation Grant No. 95-0702-01-01-3.

\bsp

\clearpage

\begin{table*}
 \centering
 \begin{minipage}{120mm}
\caption{Observing log for NGC 1798 and NGC 2192.}
\begin{tabular}{cccccc} \\
 Date & Target & Filter & Seeing & Telescope & Condition \\ [10pt]
96.11.11 & NCC 1798  & $UBV$ & $2''.3$ &
SAO\footnote{Sobaeksan Astronomical Observatory}-61cm & Photometric \\
97.01.12 & NCC 1798  & $UBVI$ & $2''.2$ & SAO-61cm & Non-photometric \\
97.02.12 & NCC 1798, NGC 2192  & $UBVI$ & $3''.3$ & SAO-61cm & Photometric \\
97.02.13 & NCC 1798, NGC 2192  & $UBVI$ & $2''.5$ & SAO-61cm & Non-photometric \\
97.03.17 & NCC 2192  & $UBVI$ & $2''.2$ & SAO-61cm & Photometric \\
97.10.21 & NCC 1798  & $BVI$ & $2''.3$ & SAO-61cm & Non-photometric \\
97.10.24 & NCC 2192  & $BVI$ & $2''.7$ & SAO-61cm & Non-photometric \\
97.03.04 & NCC 1798  & $BV$ & $2''.7$ & 
VBO\footnote{Vainu Bappu Observatory}-2.3m & Non-photometric 
\end{tabular}
\end{minipage}
\end{table*}

\begin{table}
 \centering
\caption{Transformation coefficients for the standard stars.}
\begin{tabular}{ccccccc} \\
 Date & Color & $a$ & $k$ & $Z$ & rms & n(stars) \\ [10pt]
96.11.11 & $V$ & 0.003 & --0.101 & --6.040 & 0.009 & 19 \\
         & $(B-V)$ & 1.090 & --0.155 & --0.467 & 0.013 & 17 \\
         & $(U-B)$ & 1.008 & --0.154 & --1.711 & 0.034 & 16 \\
97.02.11 & $V$ & 0.028 & --0.198 & --6.050 & 0.009 & 34 \\
         & $(B-V)$ & 1.150 & --0.118 & --0.593 & 0.010 & 35 \\
         & $(U-B)$ & 1.079 & --0.324 & --1.575 & 0.032 & 26 \\
         & $(V-I)$ & 0.983 & --0.130 & \phantom{--}0.302 & 0.008 & 29 \\
97.02.13 & $V$ & --0.007 & --0.176 & --6.005 & 0.010 & 34 \\
         & $(B-V)$ & 1.160 & --0.133 & --0.588 & 0.013 & 34 \\
         & $(U-B)$ & 1.079 & --0.349 & --1.526 & 0.023 & 20 \\
         & $(V-I)$ & 0.986 & --0.097 & \phantom{--}0.271 & 0.015 & 32 \\
97.03.17 & $V$ & -0.019 & --0.225 & --6.068 & 0.015 & 48 \\
         & $(B-V)$ & 1.221 & --0.093 & --0.781 & 0.018 & 44 \\
         & $(U-B)$ & 1.008 & --0.309 & --1.458 & 0.027 & 35 \\
         & $(V-I)$ & 0.956 & --0.157 & \phantom{--}0.314 & 0.020 & 52
\end{tabular}
\end{table}

\begin{table}
 \centering
\caption{$UBVI$ photometry of the bright stars in the C-region of NGC 1798.}
\begin{tabular}{rrrcccc} \\
ID & X[px] & Y[px] & $V$ & $(B-V)$ & $(U-B)$ & $(V-I)$ \\ [10pt]
   1 &    723.2 &   1110.8 &   15.860 &    0.779 &    0.324 &    0.914  \\ 
   3 &    680.5 &   1092.0 &   15.380 &    1.325 &    0.627 &    1.525  \\ 
   4 &    746.0 &   1160.3 &   15.905 &    0.741 &    0.220 &    0.954  \\ 
   5 &    741.8 &   1165.5 &   14.380 &    1.483 &    0.880 &    1.633  \\ 
   6 &    684.6 &   1030.1 &   15.814 &    0.864 &    0.342 &    1.042  \\ 
   7 &    745.3 &   1073.3 &   15.376 &    0.920 &    0.237 &    1.084  \\ 
   9 &    707.2 &   1038.8 &   14.777 &    1.486 &    0.723 &    1.653  \\ 
  11 &    549.4 &   1153.7 &   15.305 &    1.529 &    0.915 &    1.666  \\ 
  12 &    547.9 &   1162.7 &   14.846 &    1.627 &    0.947 &    1.764  \\ 
  13 &    602.3 &   1068.0 &   15.325 &    1.524 &    0.847 &    1.660  \\ 
  14 &    598.1 &   1080.8 &   15.776 &    1.334 &    0.575 &    1.490  \\ 
  15 &    525.8 &   1116.8 &   15.510 &    1.002 &    0.346 &    1.169  \\ 
  18 &    833.9 &   1144.2 &   14.976 &    1.494 &    0.866 &    1.578  \\ 
  21 &    981.9 &   1125.1 &   14.711 &    1.510 &    1.031 &    1.440  \\ 
  26 &    835.7 &   1098.0 &   15.706 &    1.318 &    0.468 &    1.407  \\ 
  27 &    656.5 &   1022.1 &   15.653 &    1.416 &    0.180 &    1.571  \\ 
  28 &    659.8 &    955.9 &   15.216 &    1.216 &    0.384 &    1.385  \\ 
  29 &    656.4 &    976.8 &   13.667 &    0.661 &    0.310 &    0.777  \\ 
  30 &    674.2 &    822.5 &   12.033 &    0.445 &    0.050 &    0.532  \\ 
  33 &    668.7 &    966.3 &   15.687 &    1.313 &    0.641 &    1.446  \\ 
  34 &    764.3 &    879.9 &   15.460 &    1.337 &    0.508 &    1.408  \\ 
  38 &    749.0 &   1008.1 &   15.359 &    1.322 &    0.744 &    1.426  \\ 
  39 &    774.4 &    965.9 &   15.545 &    1.304 &    0.590 &    1.402  \\ 
  43 &    571.8 &    930.6 &   12.875 &    1.970 &    1.264 &    2.252  \\ 
  45 &    512.8 &    908.7 &   15.810 &    1.348 &    0.474 &    1.485  \\ 
  46 &    642.1 &    949.5 &   15.622 &    1.329 &    0.590 &    1.477  \\ 
  48 &    732.6 &   1211.3 &   15.821 &    1.353 &    0.643 &    1.492  \\ 
  50 &    678.8 &   1210.5 &   15.297 &    1.368 &    0.770 &    1.501  \\ 
  51 &    742.1 &   1244.4 &   14.799 &    0.747 &   -0.046 &    1.216  \\ 
  52 &    673.1 &   1250.2 &   15.180 &    1.306 &    0.780 &    1.457  \\ 
  56 &    728.2 &   1183.4 &   15.665 &    1.405 &    0.803 &    1.545  \\ 
  57 &    614.9 &   1229.4 &   13.085 &    2.177 &    0.635 &    3.581  \\ 
  62 &    563.0 &   1338.4 &   15.909 &    0.706 &    0.638 &    0.778  \\ 
  65 &    913.9 &   1203.0 &   13.686 &    1.249 &    0.851 &    1.354  \\ 
  67 &    792.7 &   1314.2 &   15.601 &    0.975 &    0.470 &    1.275  \\ 
  68 &    932.6 &   1268.4 &   14.466 &    0.717 &    0.197 &    0.763  \\ 
  70 &    801.1 &   1338.7 &   15.561 &    1.327 &    0.785 &    1.442  \\ 
  72 &    827.6 &   1315.5 &   13.526 &    0.967 &    0.521 &    1.161  \\ 
  78 &    444.0 &   1094.8 &   14.605 &    1.579 &    0.825 &    1.723  \\ 
  82 &    684.3 &    822.6 &   15.828 &    0.951 &    0.128 &    1.189  \\ 
  97 &    598.5 &   1330.3 &   15.580 &    1.336 &    0.710 &    1.483   
\end{tabular}
\end{table}

\begin{table}
 \centering
\caption{$UBVI$ photometry of the bright stars in the C-region of NGC 2192.}
\begin{tabular}{rrrcccc} \\
ID & X[px] & Y[px] & $V$ & $(B-V)$ & $(U-B)$ & $(V-I)$ \\ [10pt]
   4 &    502.4 &    879.6 &   14.983 &    0.432 &    0.114 &    0.424  \\ 
   5 &    537.5 &    891.7 &   12.969 &    0.924 &    0.475 &    0.930  \\ 
   7 &    535.1 &    931.3 &   15.170 &    0.598 &   -0.201 &    0.483  \\ 
   9 &    434.8 &    947.6 &   13.676 &    0.570 &   -0.056 &    0.648  \\ 
  10 &    562.3 &    991.9 &   14.037 &    1.096 &    0.612 &    1.043  \\ 
  11 &    412.9 &    884.8 &   14.165 &    1.078 &    0.608 &    1.072  \\ 
  12 &    598.4 &    778.7 &   15.184 &    0.464 &   -0.030 &    0.477  \\ 
  13 &    595.9 &    930.0 &   14.480 &    0.502 &    0.172 &    0.549  \\ 
  14 &    625.8 &    971.8 &   14.743 &    0.720 &    0.089 &    0.741  \\ 
  16 &    614.3 &    684.2 &   14.397 &    0.766 &    0.175 &    0.758  \\ 
  17 &    463.3 &    689.5 &   14.766 &    0.466 &    0.012 &    0.480  \\ 
  18 &    553.3 &    710.7 &   15.492 &    0.447 &   -0.135 &    0.428  \\ 
  19 &    591.7 &    723.8 &   14.466 &    0.545 &    0.093 &    0.566  \\ 
  21 &    623.0 &    867.7 &   12.634 &    1.102 &    0.713 &    1.108  \\ 
  22 &    648.0 &    929.0 &   15.338 &    0.394 &    0.115 &    0.399  \\ 
  26 &    461.1 &    711.3 &   14.162 &    1.086 &    0.670 &    1.091  \\ 
  29 &    583.9 &    802.3 &   15.094 &    0.600 &   -0.016 &    0.644  \\ 
  32 &    720.4 &    889.5 &   15.326 &    0.370 &    0.071 &    0.410  \\ 
  37 &    395.9 &    778.3 &   15.264 &    0.396 &    0.031 &    0.379  \\ 
  38 &    405.5 &    778.5 &   14.890 &    0.481 &    0.101 &    0.409  \\ 
  40 &    258.2 &   1042.9 &   15.431 &    0.424 &    0.120 &    0.484  \\ 
  42 &    376.6 &    883.9 &   15.187 &    0.455 &    0.136 &    0.470  \\ 
  44 &    180.1 &    908.7 &   14.320 &    0.287 &    0.112 &    0.295  \\ 
  47 &    367.1 &    938.3 &   14.897 &    0.403 &    0.252 &    0.459  \\ 
  51 &    183.3 &    998.7 &   14.202 &    1.114 &    0.552 &    1.108  \\ 
  55 &    475.8 &   1052.5 &   14.637 &    0.388 &    0.167 &    0.465  \\ 
  56 &    389.1 &   1055.1 &   13.606 &    0.652 &    0.256 &    0.777  \\ 
  58 &    481.4 &   1074.1 &   14.347 &    0.450 &    0.115 &    0.520  \\ 
  59 &    514.9 &   1085.1 &   15.097 &    0.429 &    0.137 &    0.517  \\ 
  60 &    478.5 &   1089.6 &   14.236 &    0.528 &    0.159 &    0.582  \\ 
  61 &    498.5 &   1127.9 &   14.609 &    0.509 &    0.135 &    0.578  \\ 
  63 &    278.7 &   1154.4 &   14.372 &    1.070 &    0.599 &    1.053  \\ 
  64 &    527.4 &   1158.7 &   15.418 &    0.446 &    0.245 &    0.510  \\ 
  65 &    343.4 &   1167.8 &   14.266 &    1.012 &    0.618 &    1.035  \\ 
  66 &    362.2 &   1169.3 &   14.266 &    0.879 &    0.469 &    0.879  \\ 
  74 &    346.9 &   1039.8 &   14.581 &    0.452 &    0.124 &    0.510  \\ 
  77 &    688.9 &   1124.2 &   15.420 &    0.390 &    0.290 &    0.400  \\ 
  83 &    678.9 &   1079.7 &   15.495 &    0.464 &    0.107 &    0.504  \\ 
  84 &    624.2 &   1042.8 &   13.501 &    0.385 &    0.217 &    0.389  \\ 
  88 &    579.6 &   1028.9 &   14.825 &    0.456 &    0.162 &    0.530   
\end{tabular}
\end{table}

\begin{table}
 \centering
\caption{Basic properties of NGC 1798 and NGC 2192.}
\begin{tabular}{lcc} \\
  Parameter & NGC 1798  & NGC 2192 \\
       &  & \\ [10pt]
RA(2000) & 5$^h$ 11$^m$ 40$^s$ & 6$^h$ 15$^m$ 11$^s$\\
DEC(2000) & 47$^\circ$ 40$'$ 37$''$ &  39$^\circ$ 51$'$ 1$''$\\
$l$ & 160$^\circ$.76 & 173$^\circ$.41 \\
$b$ & 4$^\circ$.85 & 10$^\circ$.64 \\
Age & $1.4\pm0.2$ Gyrs & $1.1\pm0.1$ Gyrs\\
$E(B-V)$ & $0.51\pm0.04$ & $0.20\pm0.04$\\
{\rm [Fe/H]} & $-0.47\pm0.15$ dex & $-0.31\pm0.15$ dex\\
$(m-M)_0$ & $13.1\pm0.2$ & $12.7\pm0.2$\\
distance & $4.2\pm0.3$ kpc & $3.5\pm0.3$ kpc\\
$R_{GC}$ & 12.5 kpc &  11.9 kpc\\
z & 355 pc & 646 pc\\
diameter & 10.2 pc ($8'.3$) &  7.5 pc ($7'.3$) 
\end{tabular}
\end{table}

\clearpage

\leftline{\bf FIGURE CAPTIONS}

\bigskip
\noindent {\bf Figure 1.} 
A grey scale map of the $V$ CCD image of NGC 1798 made by mosaicing the images
of observed regions. North is right and east is up. The size of the field is $10'.6\times 13'.8$.
The circle and lines represent the boundary of each region described in the text.
\medskip

\noindent {\bf Figure 2.} 
A grey scale map of the $V$ CCD image of NGC 2192 made by mosaicing the images
of observed regions. North is right and east is up. The size of the field is $7'.2\times 10'.8$.
The circle and line represent the boundary of each region described in the text.
\medskip


\noindent {\bf Figure 3.} 
Projected surface number density of the stars with $V<19.5$ mag
in the NGC 1798 area.
\medskip

\noindent {\bf Figure 4.} 
$V$--($B$--$V$) color-magnitude diagrams of NGC 1798: the entire region,
the C-region, the N-region, and Fb-region.
The square represents the position of the red giant clump.
\medskip

\noindent {\bf Figure 5.}  
$V$--($V$--$I$) color-magnitude diagrams of NGC 1798: the entire region,
the C-region, the N-region, and the Fir-region plus Fi-region.
The square represents the position of the red giant clump.
\medskip

\noindent {\bf Figure 6.}  
($U$--$B$)--($B$--$V$) diagram of stars with small photometric errors in 
the C-region of NGC 1798.
The dotted line and solid line represent, respectively,
 the mean line for the giants with solar abundance (III) 
given by Schmidt-Kaler (1982) and the mean line for the Padova isochrones
with [Fe/H] = --0.47, which were shifted according to the reddening
of $E(B-V)=0.55$.
\medskip

\noindent {\bf Figure 7.}  
ZAMS fitting for the C-region of NGC 1798.
The solid line represents the empirical ZAMS for [Fe/H] = --0.47, 
and $Y=0.28$, shifted according to the reddening and distance of NGC 1798.
The dashed lines represent the upper and lower envelope corresponding to
the fitting errors. 
\medskip

\noindent {\bf Figure 8.}  
Isochrone fitting for the C-region of NGC 1798 in the color-magnitude diagrams.
The solid line represents the Padova isochrone for age = 1.4 Gyrs, 
[Fe/H] = --0.47, shifted according to the reddening and distance of NGC 1798.
The dashed lines represents isochrones for ages of 1.2 and 1.6 Gyrs. 
\medskip

\noindent {\bf Figure 9.}  
(a) Luminosity functions of the main sequence stars in the C-region
 (filled circles), N-region (open squares) and R-region plus Fir-region 
(open triangles) of NGC 1798. 
(b) The luminosity function for the C-region based on a smaller 
bin size of 0.2 mag.
\medskip

\noindent {\bf Figure 10.}  
Projected surface number density of the stars with $V<18$ mag 
in the NGC 2192 area.
\medskip

\noindent {\bf Figure 11.}  
$V$--($B$--$V$) color-magnitude diagrams of NGC 2192: the entire region,
the C-region, the N-region, and the F-region.
The square represents the position of the red giant clump.
\medskip

\noindent {\bf Figure 12.}  
$V$--($V$--$I$) color-magnitude diagrams of NGC 2192: the entire region,
the C-region, the N-region, and the F-region.
The square represents the position of the red giant clump.
\medskip

\noindent {\bf Figure 13.}  
($U$--$B$)--($B$--$V$) diagram of stars with small photometric errors in 
the C-region of NGC 2192.
The dotted line and solid line represent, respectively,
 the mean line for the giants with solar abundance (III) 
given by Schmidt-Kaler (1982) and the mean line for the Padova isochrones
with [Fe/H] = --0.31, which were shifted according to the reddening
of $E(B-V)=0.19$.
\medskip

\noindent {\bf Figure 14.}  
ZAMS fitting for the C-region of NGC 2192.
The solid line represents the empirical ZAMS for [Fe/H] = --0.31, 
and $Y=0.28$, shifted according to the reddening and distance of NGC 2192.
The dashed lines represent the upper and lower envelope corresponding to
the fitting errors. 
\medskip

\noindent {\bf Figure 15.}  
isochrone fitting for the C-region of NGC 2192 in the color-magnitude diagrams.
The solid line represents the Padova isochrone for age = 1.1 Gyrs, 
[Fe/H] = --0.31, shifted according to the reddening and distance of NGC 2192.
The dashed lines represents isochrones for ages of 1.0 and 1.2 Gyrs. 
\medskip

\noindent {\bf Figure 16.}  
(a) Luminosity functions of the main sequence stars in the C-region
 (filled circles) and N-region (open squares) of NGC 2192. 
(b) Comparison of the luminosity functions of NGC 2192
 (filled circles), NGC 1798 (open triangles) and NGC 7789 (open circles). 
\medskip

\noindent {\bf Figure 17.}  
Metallicity versus the galactocentric distance of NGC 1798 
(the filled circle) and NGC 2192 (the filled square)
compared with other old open clusters (open circles).
\medskip

\label{lastpage}

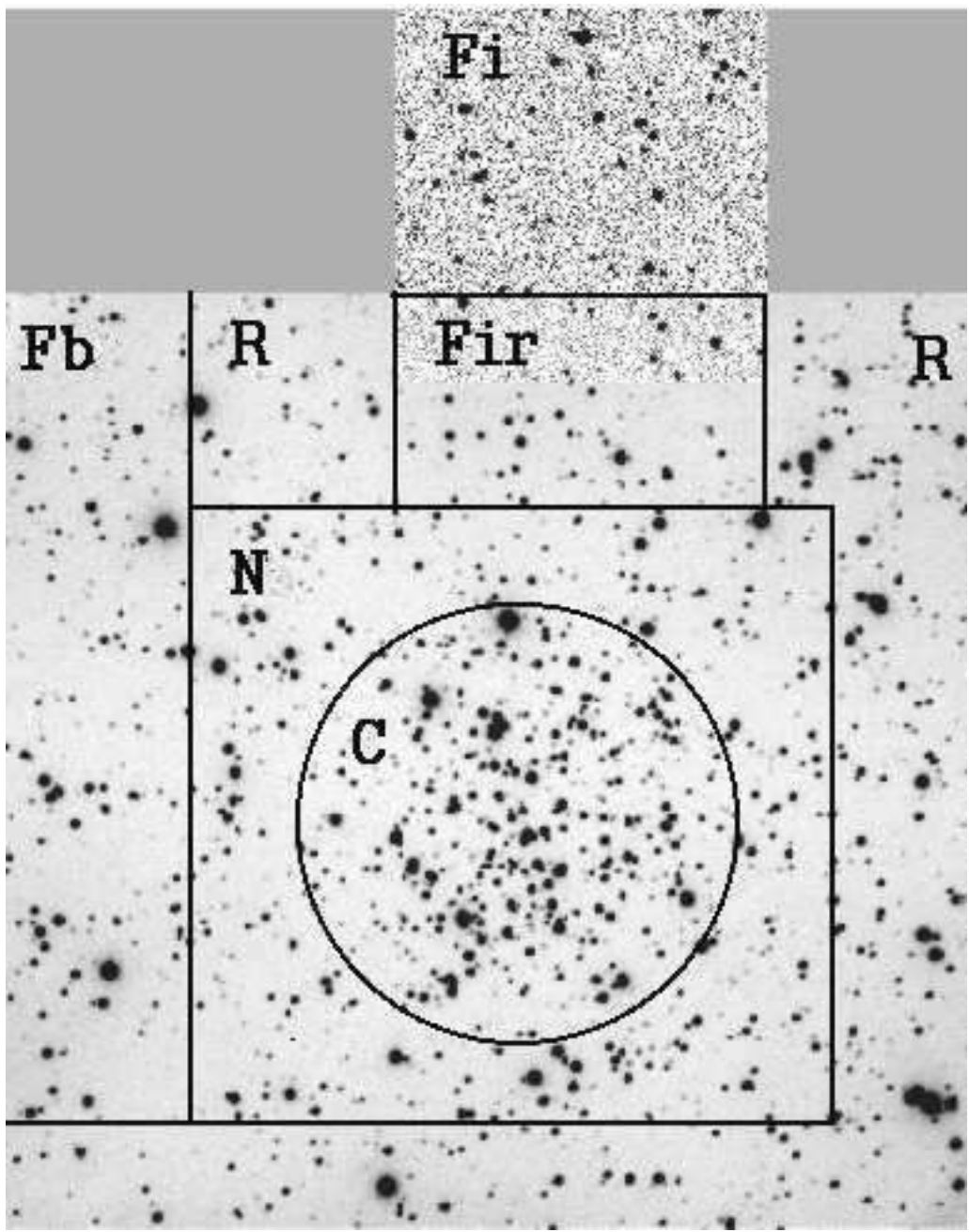
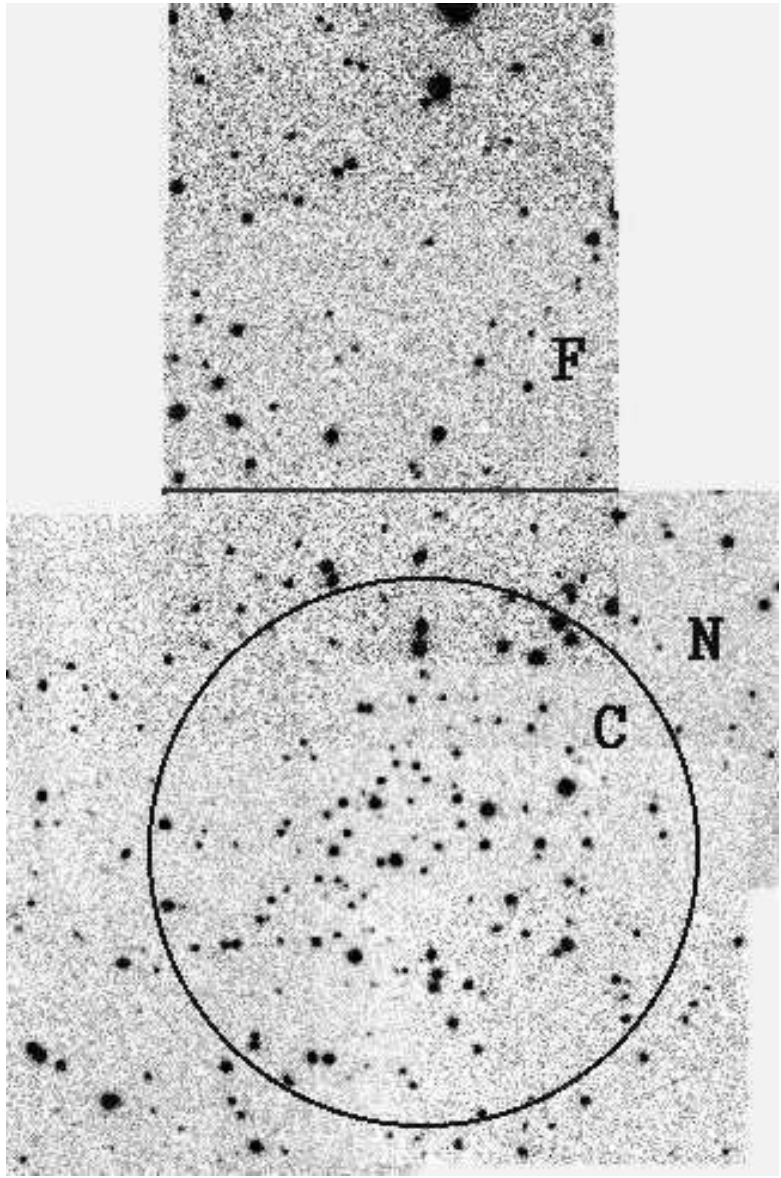
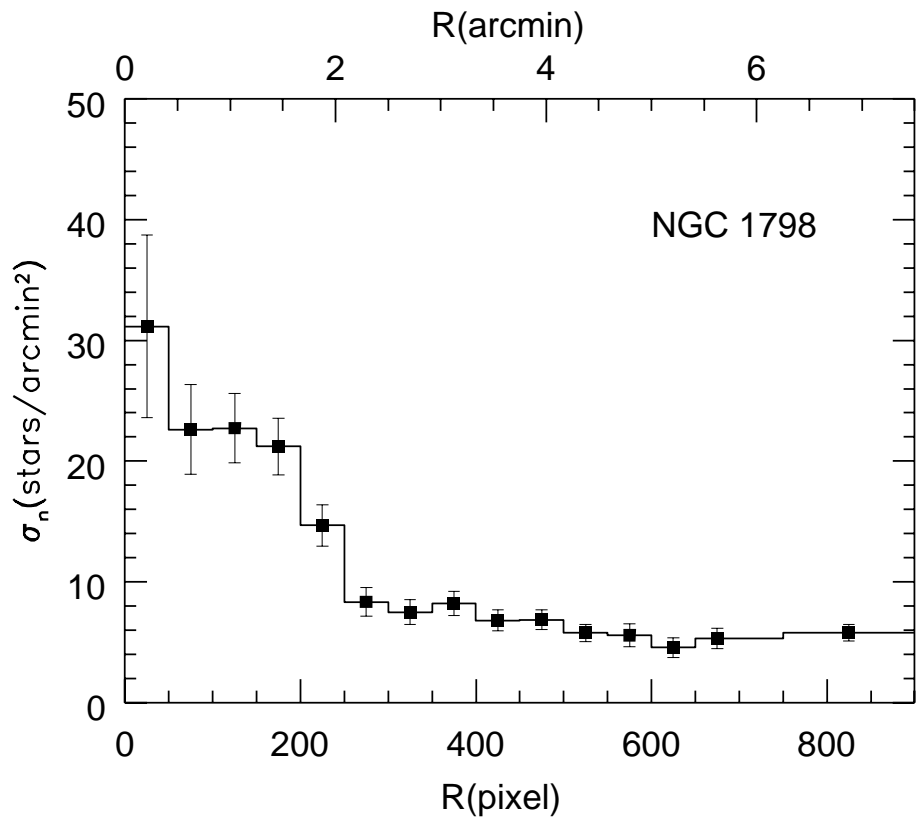
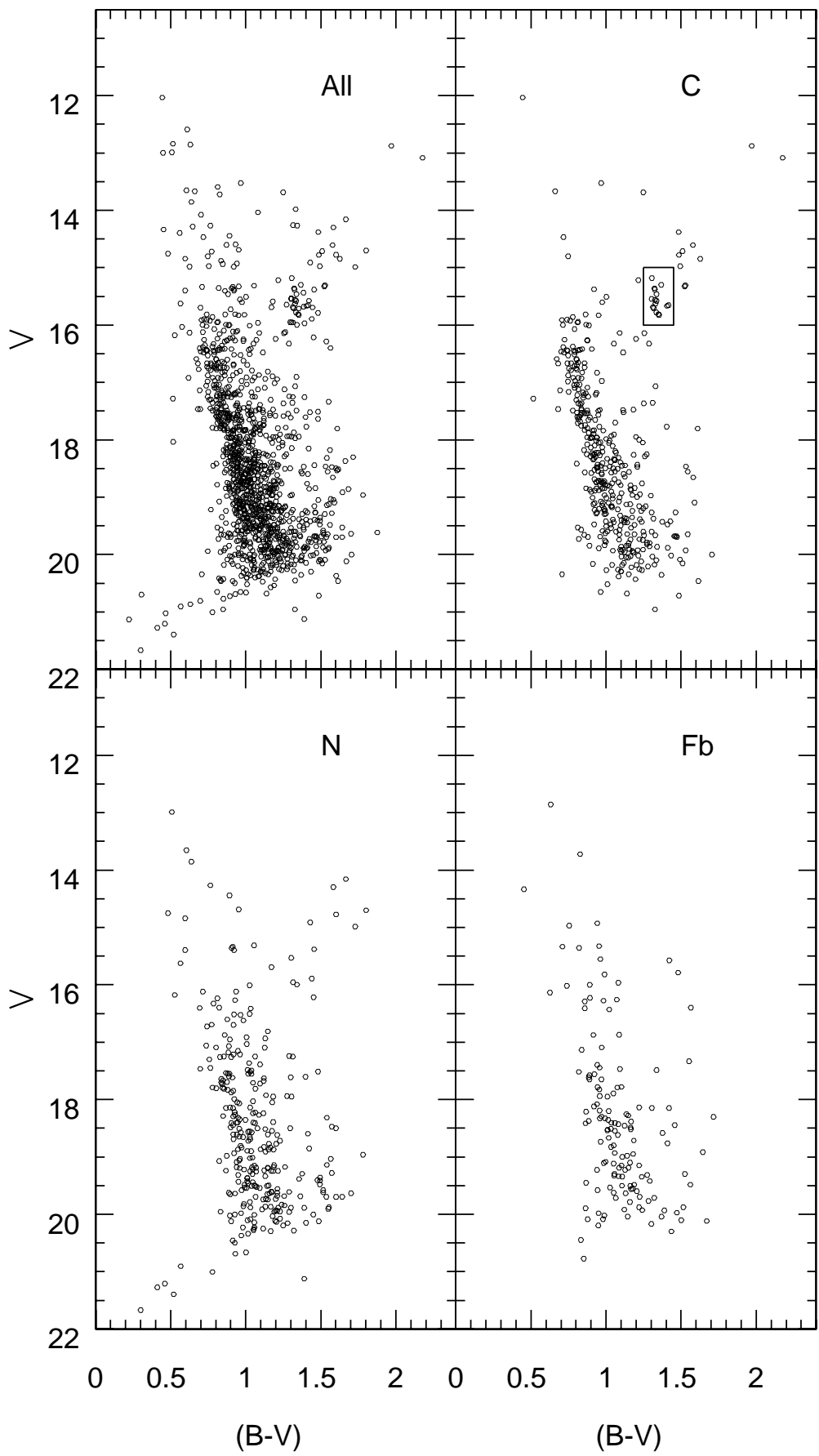
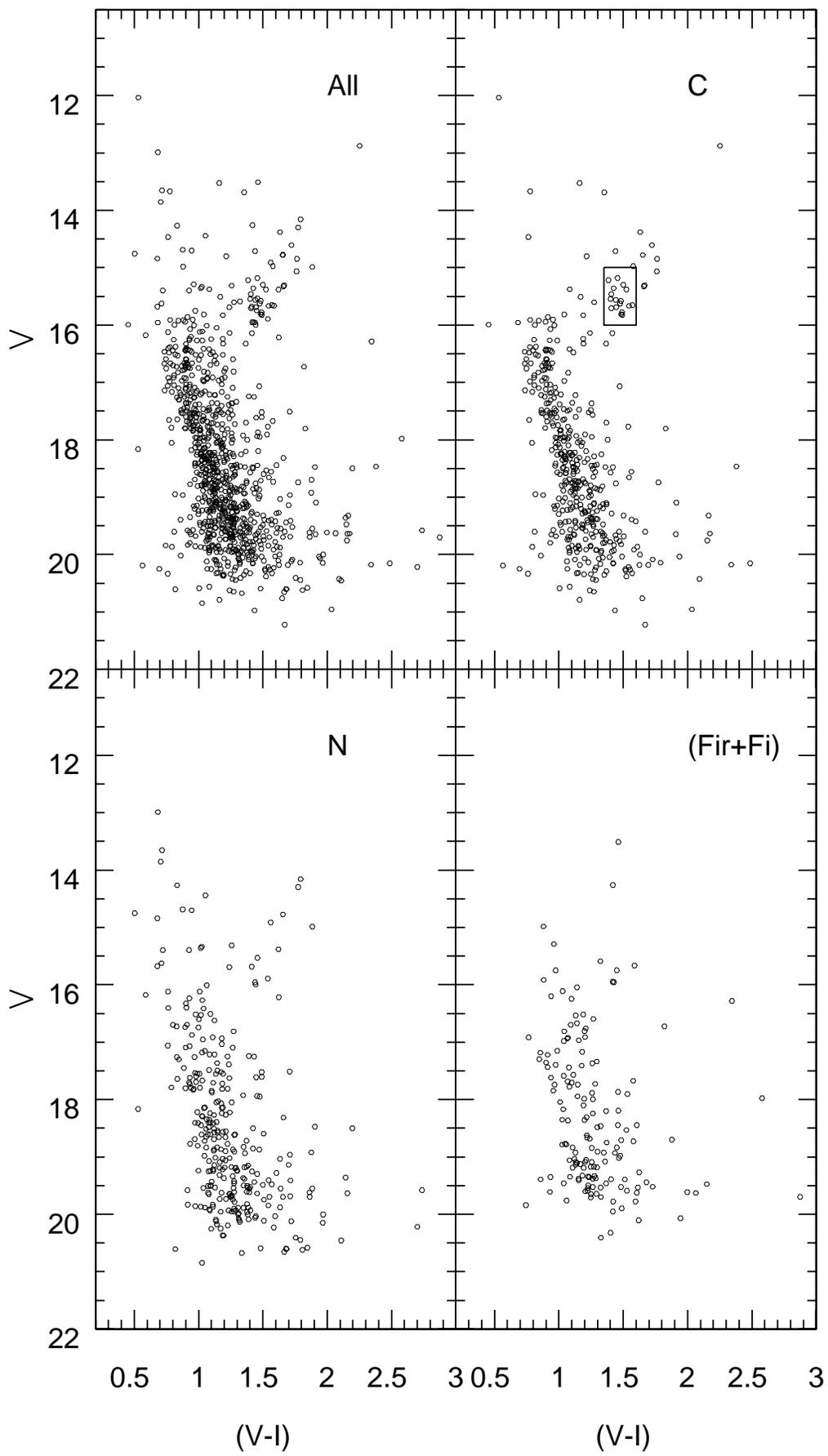
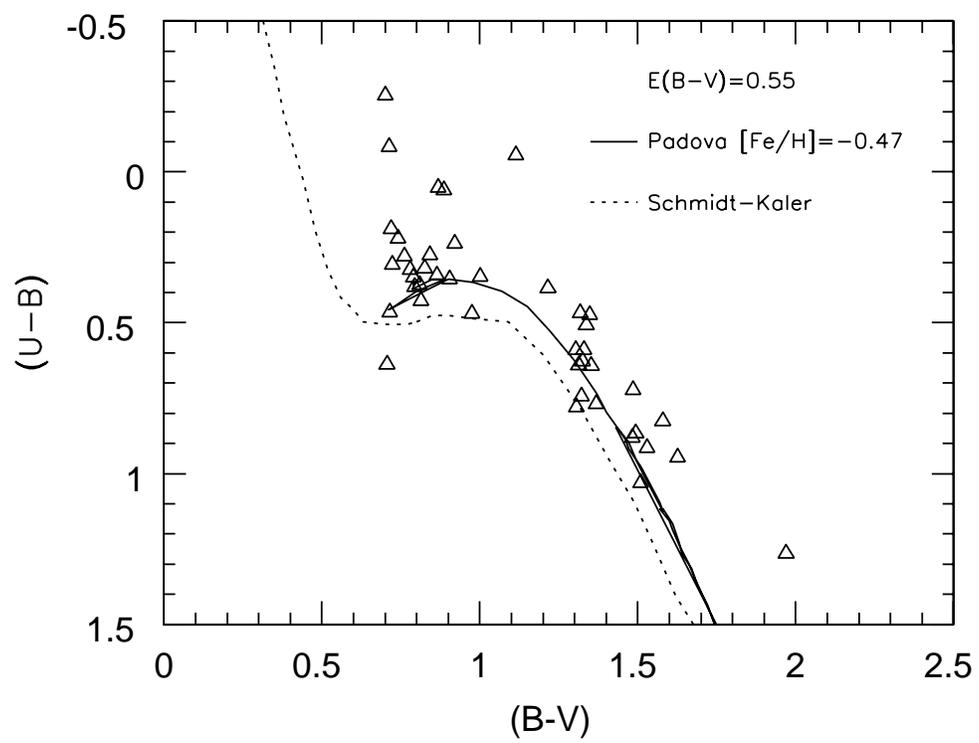
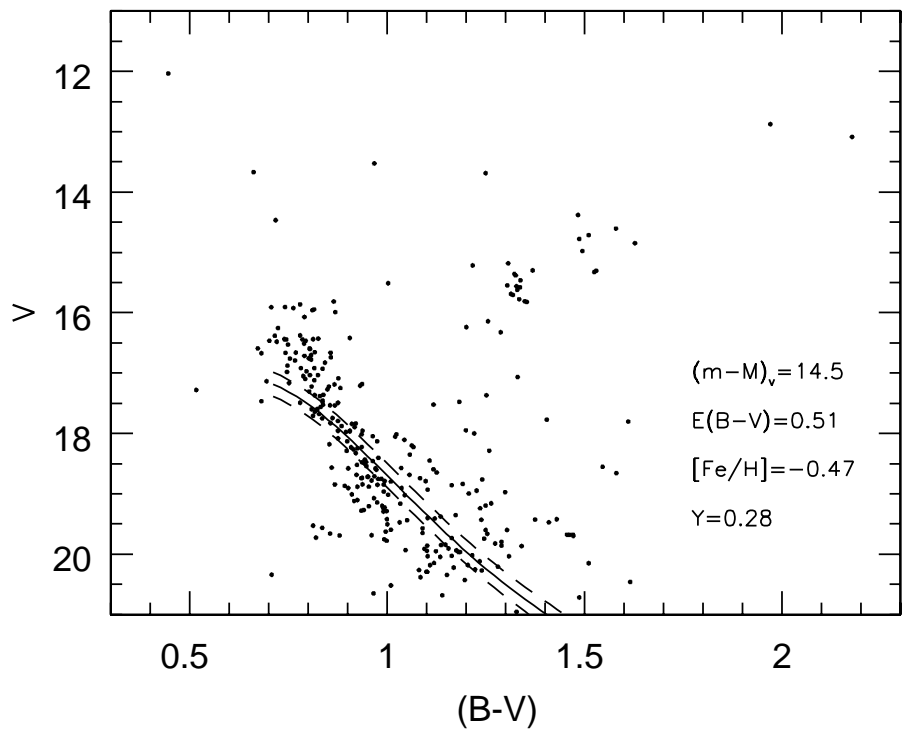
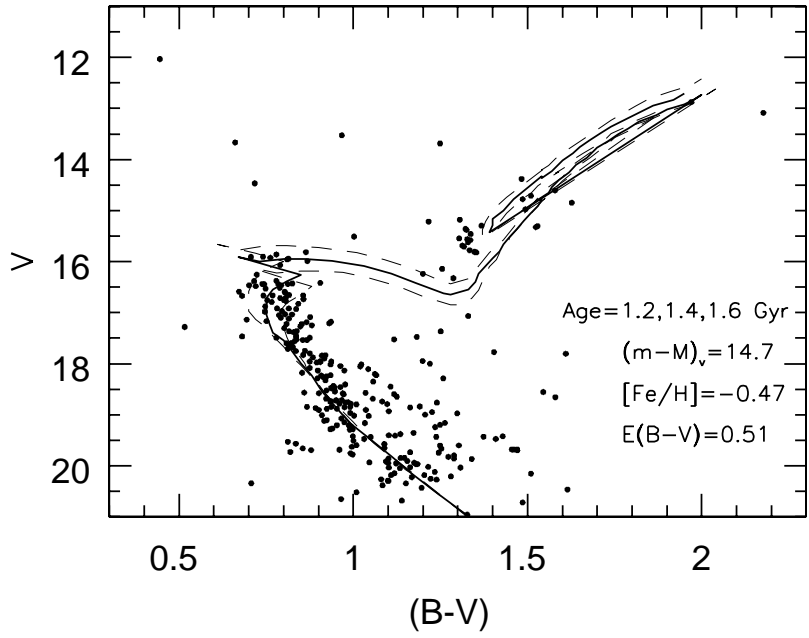
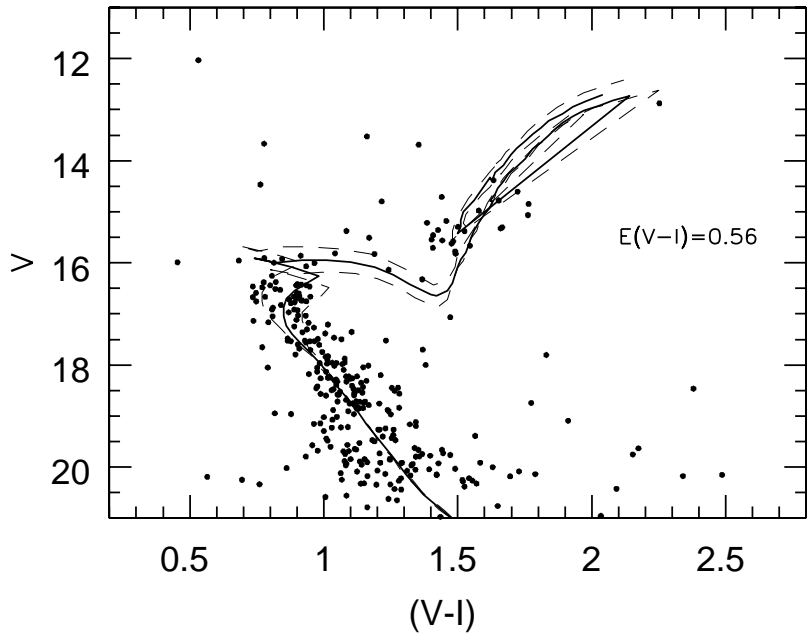
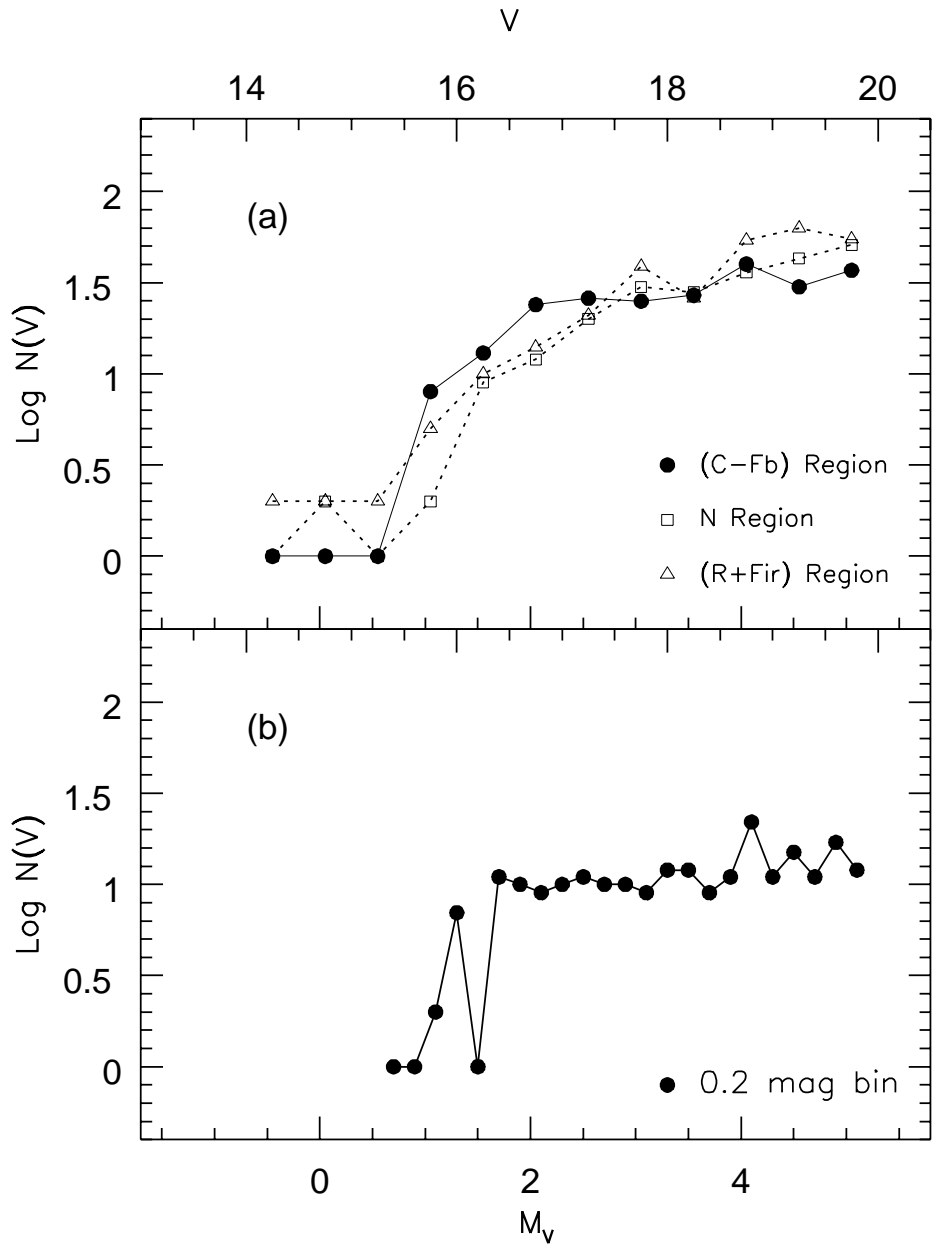
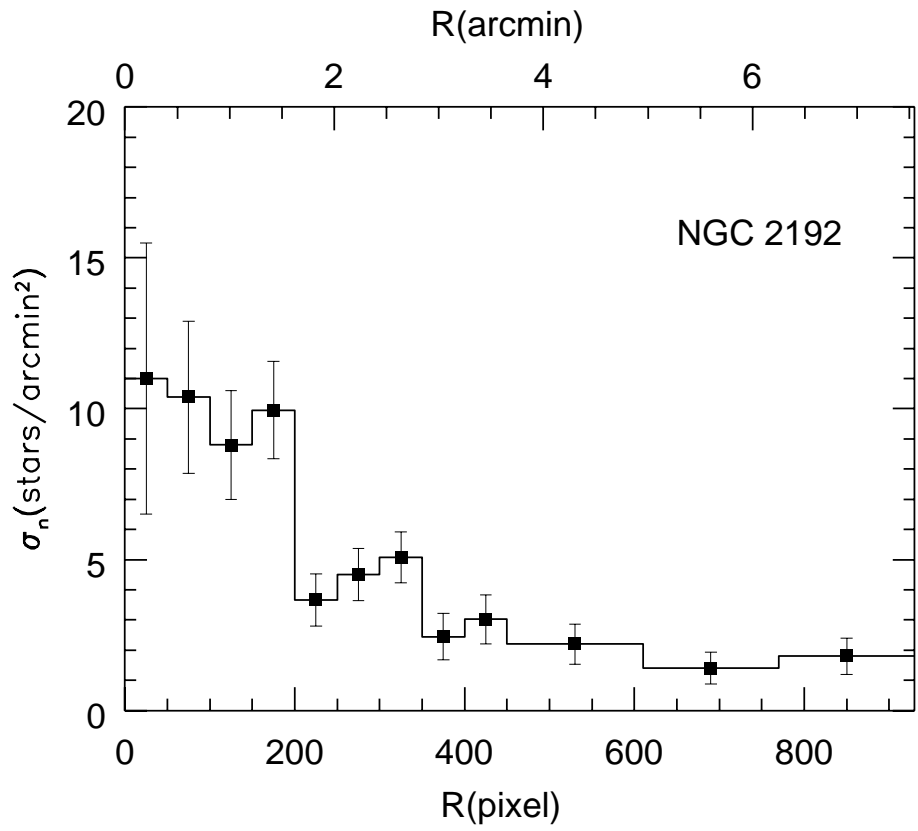
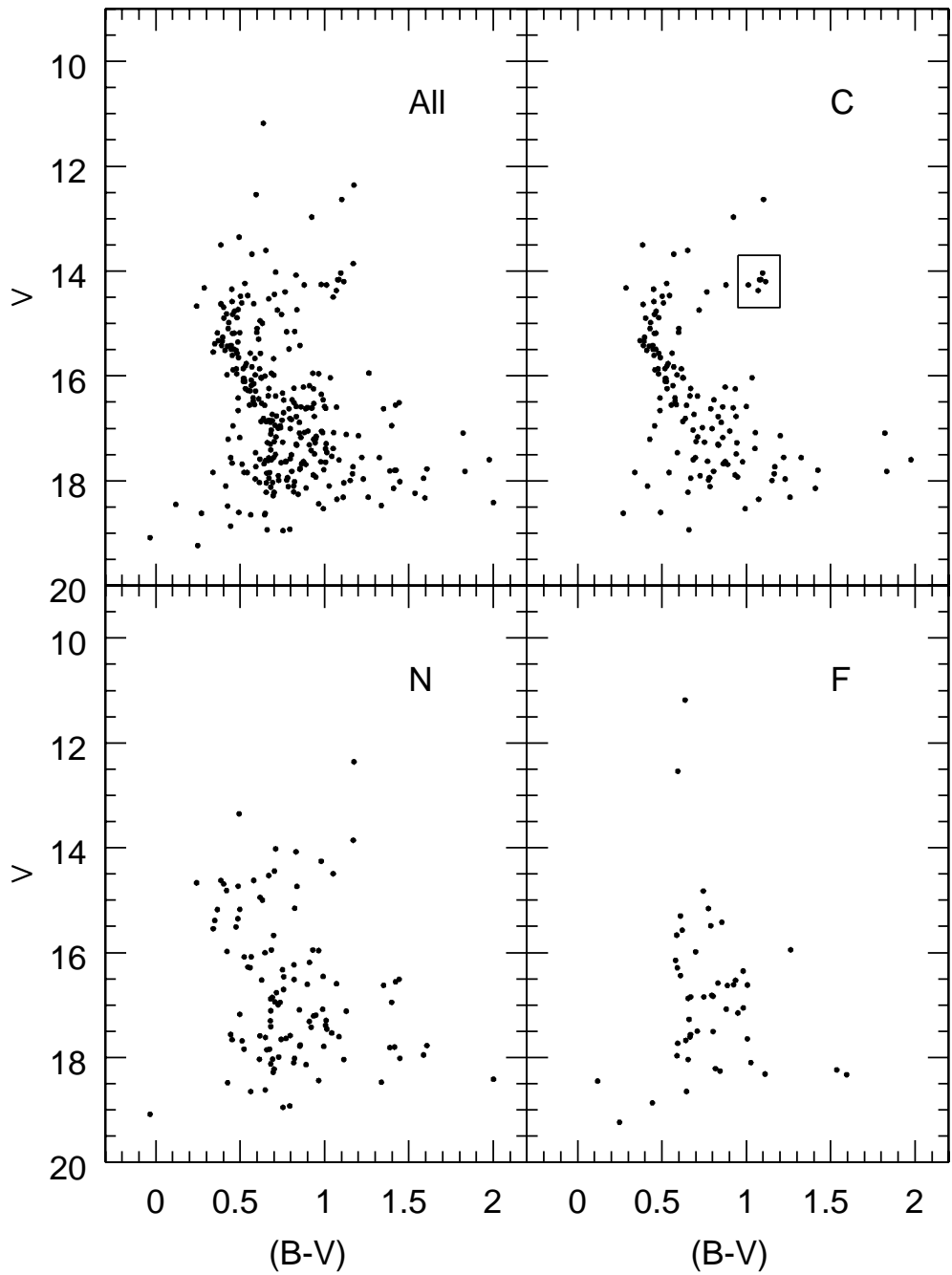
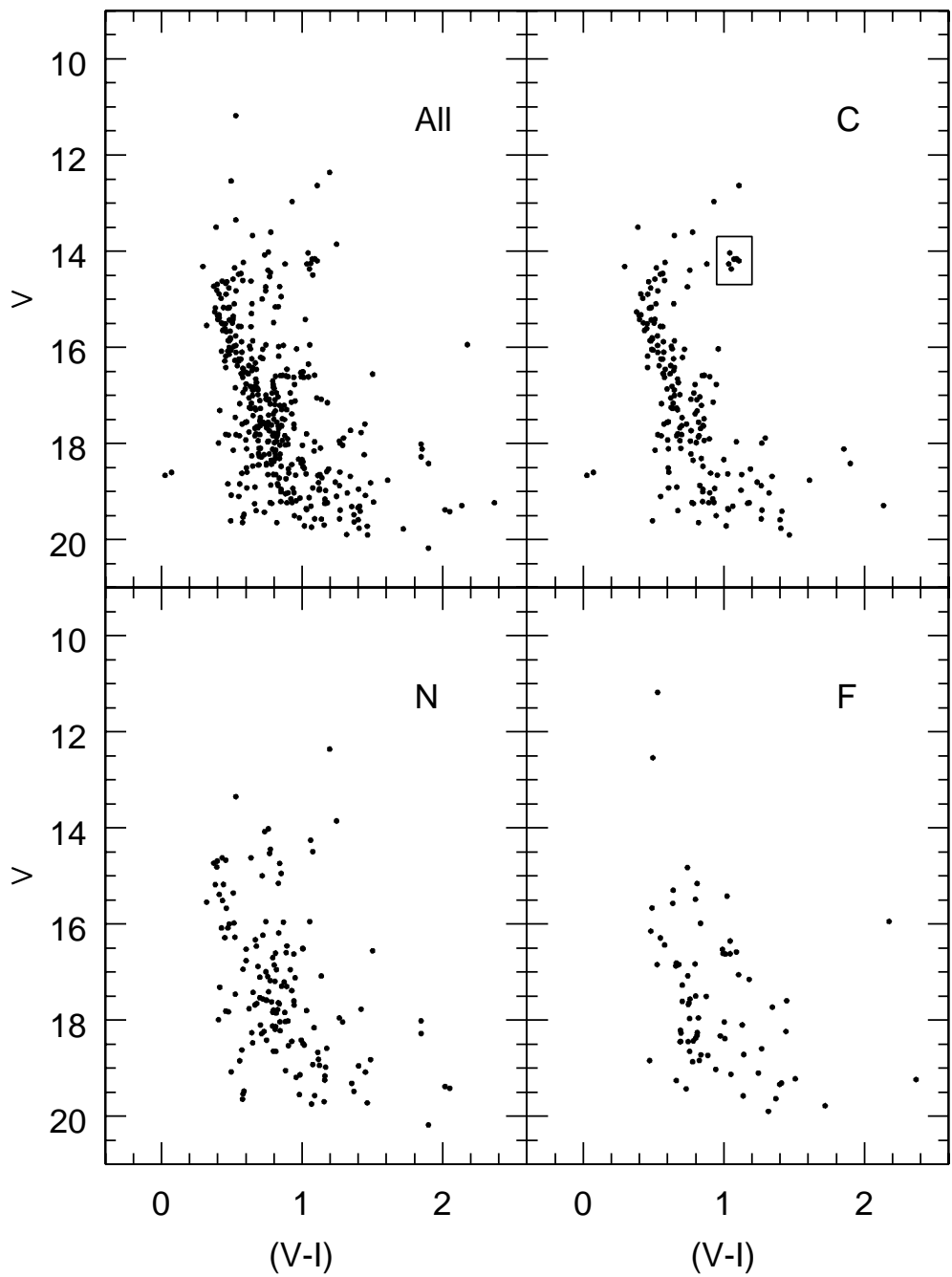
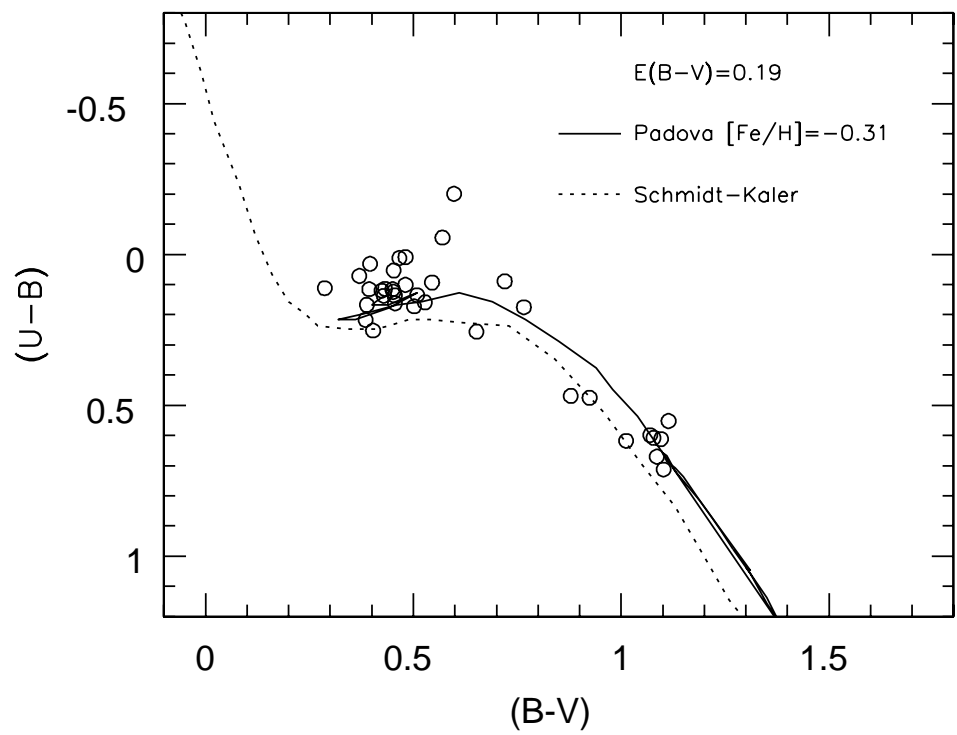
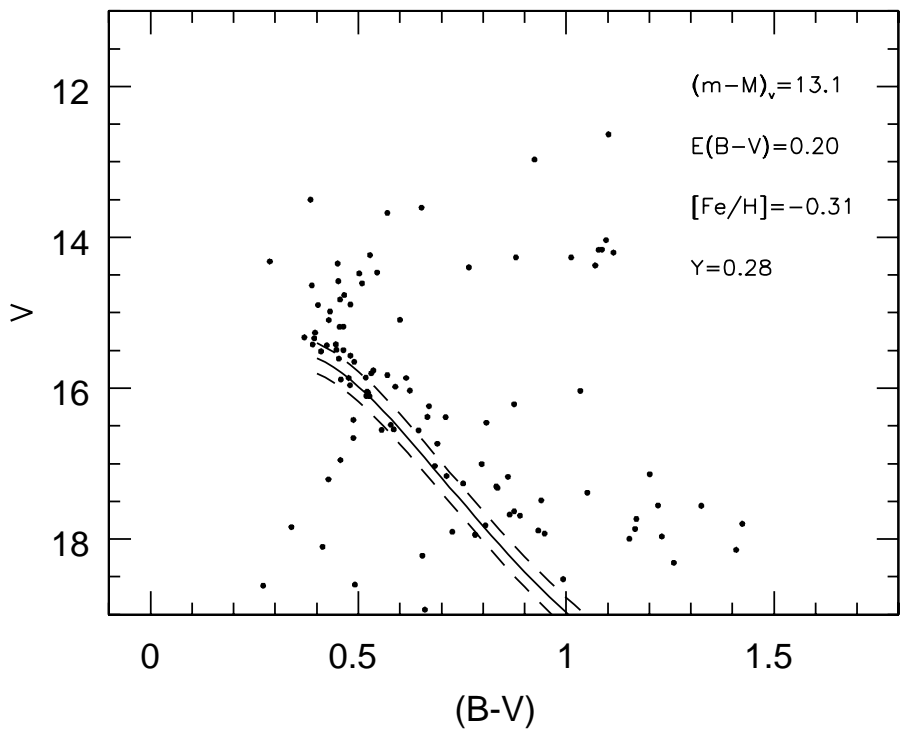
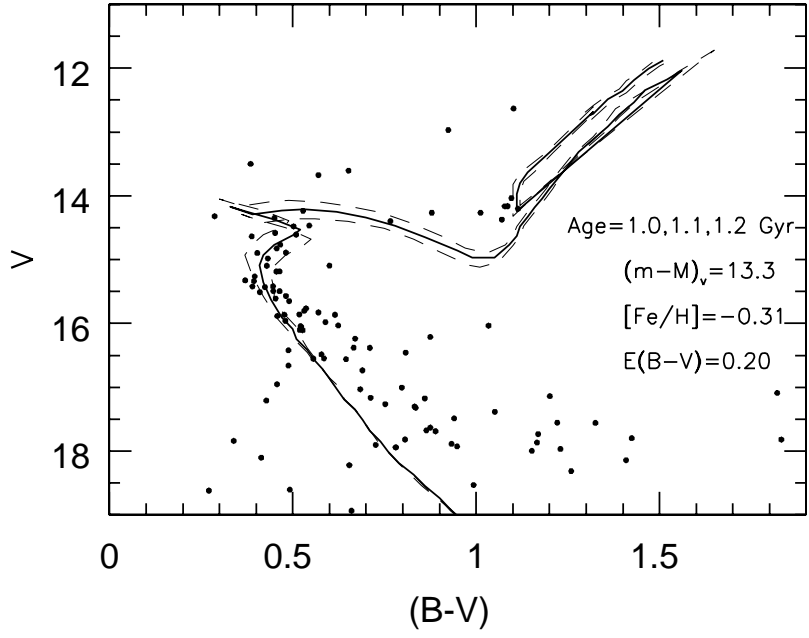
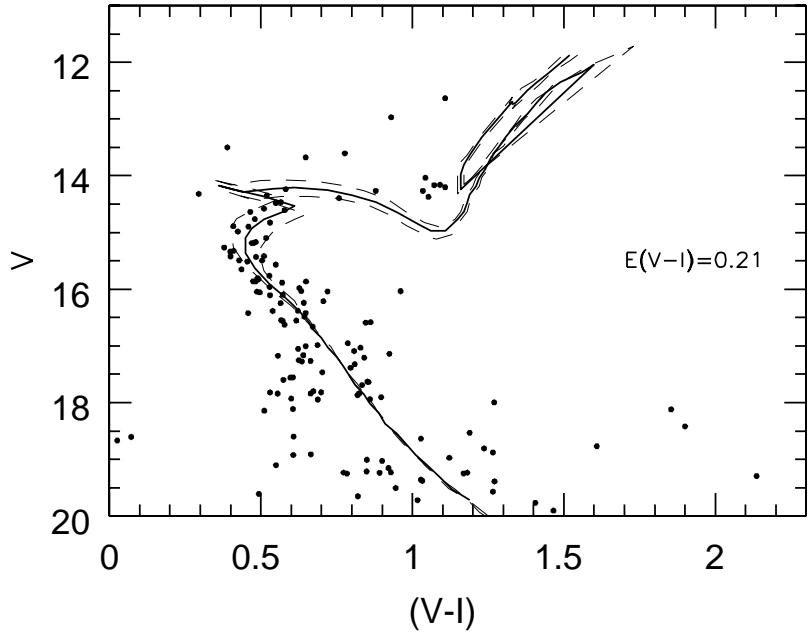
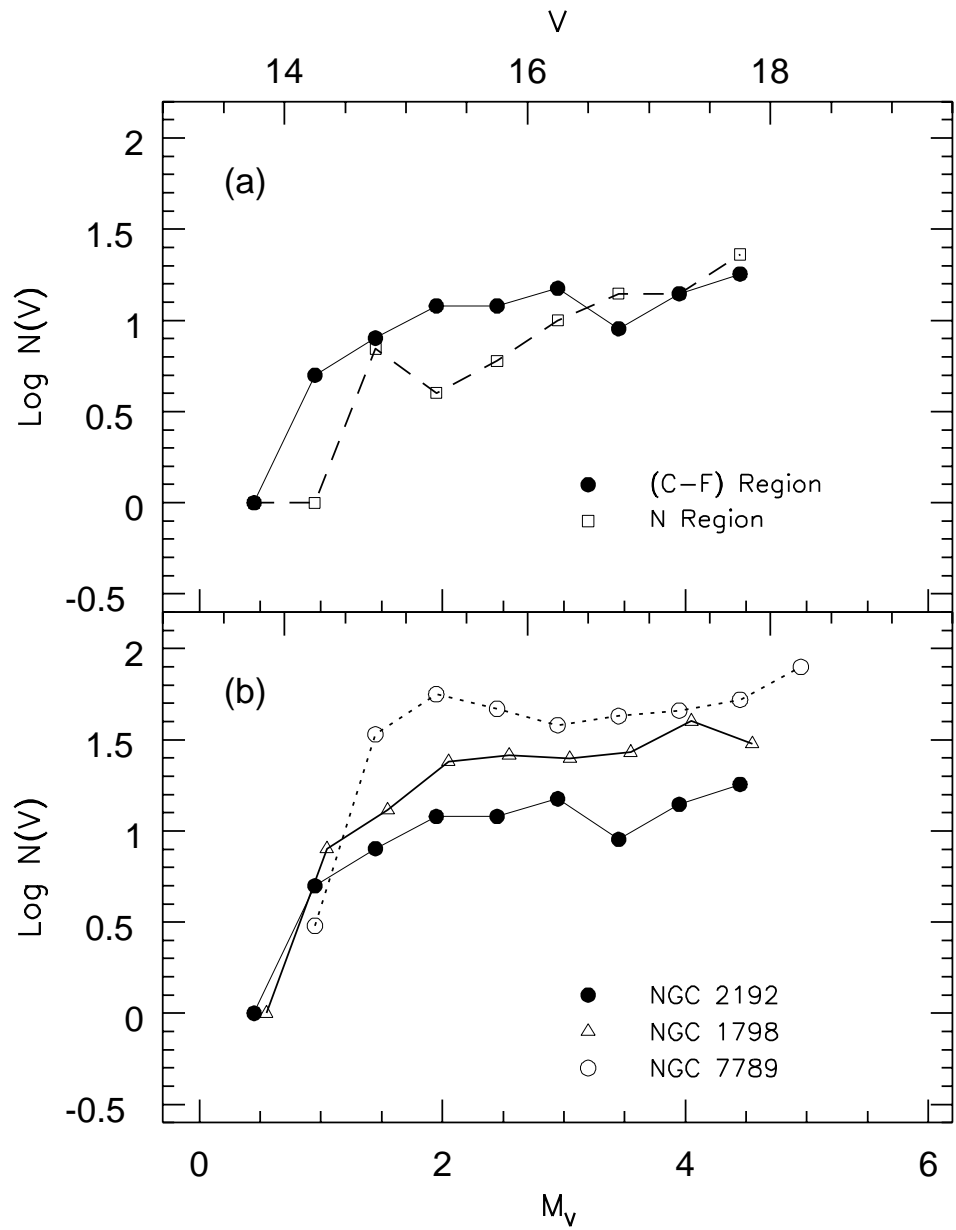
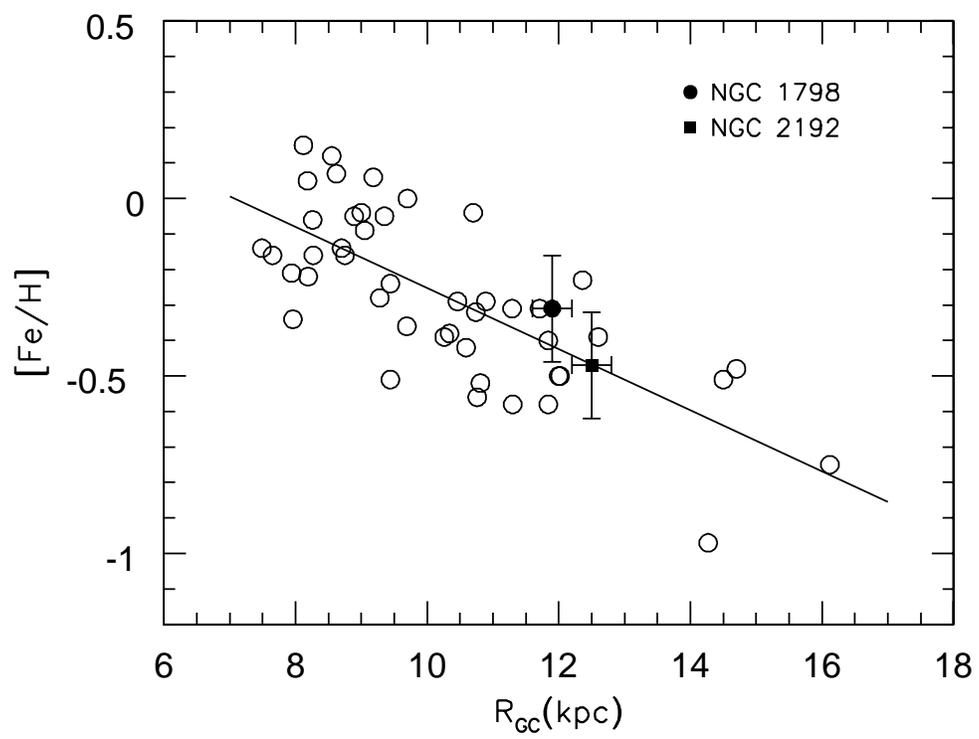
\end{document}